# Zircon to monazite phase transition in CeVO$_4$

V. Panchal[1], S. López-Moreno[2], D. Santamaría-Pérez[3], D. Errandonea[1,†], F.J. Manjón[4],

P. Rodríguez-Hernandez[2], A. Muñoz[2], S. N. Achary[5], and A. K. Tyagi[5]

[1] Departamento de Física Aplicada-ICMUV, MALTA Consolider Team, Universidad de Valencia, Edificio de Investigación, C/Dr. Moliner 50, 46100 Burjassot (Valencia), Spain

[2] Departamento de Física Fundamental II, Instituto de Materiales y Nanotecnología, Universidad de La Laguna, La Laguna 38205, Tenerife, Spain

[3] Departamento de Química Física I–MALTA Consolider Team, Universidad Complutense de Madrid, Avda. Complutense s/n, 28040 Madrid, Spain

[4] Instituto de Diseño para la Fabricación y Producción Automatizada, MALTA Consolider Team, Universidad Politécnica de Valencia, Camino de Vera s/n, 46022 Valencia, Spain

[5] Chemistry Division, Bhabha Atomic Research Centre, Trombay, Mumbai 400085, India

**Abstract:** X-ray diffraction and Raman-scattering measurements on cerium vanadate have been performed up to 12 and 16 GPa, respectively. Experiments reveal that at 5.3 GPa the onset of a pressure-induced irreversible phase transition from the zircon to the monazite structure. Beyond this pressure, diffraction peaks and Raman-active modes of the monazite phase are measured. The zircon to monazite transition in CeVO$_4$ is distinctive among the other rare-earth orthovanadates. We also observed softening of external translational T(E$_g$) and internal ν$_2$(B$_{2g}$) bending modes. We attributed it to mechanical instabilities of zircon phase against the pressure-induced distortion. We additionally report lattice-dynamical and total-energy calculations which are in agreement with the experimental results. Finally, the effect of non-hydrostatic stresses on the structural sequence is studied and the equations of state of different phases are reported.

## I. Introduction

Recently significant research has been carried out in the field of zircon-type orthovanades (AVO$_4$, A = trivalent atom) due to their wide practical applications. Orthovanadates are well known for birefringent materials and optical polarizers. These materials are potential candidates for laser-host materials and also find applications in other fields such as cathodoluminescent, thermophosphors, scintillators, phosphors, and nuclear-waste storage materials [1-3]. In general, the AVO$_4$ orthovanadaes crystallizes in two polymorphs, a tetragonal zircon-type structure [space group (SG): *I4$_1$/amd*] [4] and a monoclinic monazite-type (space group: *P2$_1$/n*) [5]. Zircon structure is composed of alternating edge-sharing AO$_8$ dodecahedra and VO$_4$ tetrahedra forming chains parallel to the *c*-axis, while in the monazite structure AO$_9$ polyhedra are edge-shared with VO$_4$ tetrahedra along the *c*-axis (see Fig. 1).

The phase stability of lanthanide-based ABO$_4$ compounds depends upon the A/B cation size ratio, and those with large ionic radius, like LaVO$_4$, crystallize in monazite structure. The compound CeVO$_4$, with Ce having smaller ionic radius than La, crystallizes in the zircon structure, although it is located close to the boundary of zircon and monazite structures. Hence, the zircon phase is expected to undergo a structural phase transition at a relatively low pressure compared to other orthovanadates. In view of the nature of the structural phase transitions in these compounds it is quite imperative to understand the behavior of the zircon structured orthovanadates under hydrostatic compression and verify whether the cation A radius plays any vital role in the structural phase transitions. As the matter of fact, most of the rare-earth orthovanadates have been observed to undergo irreversible zircon to scheelite phase transition [6]. However, rare-earth-based zircon compounds with relatively large ionic radii, like CeVO$_4$, NdVO$_4$ and PrVO$_4$, have not been investigated so far.

Recent high-pressure (HP) measurements performed in orthophosphates evidence the importance of the cation A ionic radius in the sequence of structural phase transitions. Raman spectroscopy measurements and lattice-dynamic calculations supported by x-ray diffraction (XRD) in $TbPO_4$ indicate a zircon to monoclinic phase transition at 9.5 GPa [7, 8]. Similarly an XRD investigation in orthophosphates, viz. $YPO_4$ and $ErPO_4$, reports a zircon to monazite phase transition in both compounds [9]. On the other hand, $ScPO_4$, $YbPO_4$ and $LuPO_4$ undergo reversible zircon to scheelite phase transition [10, 11]. Thus, it is a key issue to find what factor governs both zircon to scheelite and zircon to monazite phase transitions in orthovanadates and orthophosphates. In the present investigation we report XRD and Raman scattering measurements in $CeVO_4$, the mineral wakefieldite-(Ce), up to 12.0 and 15.9 GPa, respectively, together with *ab initio* calculations.

## II. Experimental details

$CeVO_4$ samples were prepared by solid-state reaction of appropriate amounts of predried $Ce_2O_3$ (Indian Rare Earth Ltd. 99%) and $V_2O_5$ (Alfa-Aesar 99%). Homogeneous mixtures of the reactants were pelletized and heated at 800°C for 24 h and then cooled to ambient temperature. Further, the pellets were reground and heated again at 1100°C for 24 h. The sample obtained was characterized by XRD as a single phase of $CeVO_4$ of zircon-type structure.

Angle-dispersive powder XRD measurements both at ambient and at HP were recorded on the 135 mm Atlas CCD detector placed at 110 mm from the sample of a Xcalibur diffractometer (Oxford Diffraction Limited) using $K_{\alpha 1}$: $K_{\alpha 2}$ molybdenum radiation. The X-ray beam was collimated to a diameter of 300 μm. The same set-up was used previously to successfully characterize the HP phases in $ABO_4$ oxides up to 20 GPa [12]. On the other hand, Raman measurements both at ambient and HP were performed in the backscattering geometry using 632.8 nm HeNe laser and a Horiba Jobin-Yvon LabRAM high-resolution microspectrometer in combination with a thermoelectric-cooled multichannel charged-coupled device (CCD) detector with spectral resolution below 2 cm$^{-1}$.

For HP measurements on CeVO$_4$, finely grounded powder sample of CeVO$_4$, along with 2-μm diameter ruby balls, was loaded in a pre-indented steel gasket with a 200-μm diameter hole inside a diamond-anvil cell (DAC). A 4:1 methanol-ethanol mixture was used as pressure-transmitting medium [13, 14]. The pressure was determined using the ruby-fluorescence technique [15]. A modified Merrill-Bassett DAC was used for XRD measurements and a membrane-type DAC was used for Raman measurements.

We have also performed compression measurements in CeVO$_4$ at ambient and high-temperature using steel-belted Bridgman-type opposed tungsten-carbide anvils with a tip of 15 mm in diameter [16]. The sample was contained in a pyrophyllite chamber which consists of two pyrophyllite gaskets of optimized thickness (0.5 mm each) in split gasket geometry. Cubic boron nitride was the pressure medium used in these experiments. To increase the temperature we used a graphite heater [17]. Temperature was measured using a steel shielded K-type thermocouple and pressure by the calibration of the load applied to the anvils against HP resistivity transitions in Bi, Yb, CdTe and n-type InSe [18]. Effects of pressure in the thermocouple were neglected. Recovered samples from different experimental runs were analyzed by XRD and Raman spectroscopy.

**III. Theoretical method and computational details**

It is well known that *ab initio* methods have allowed detailed studies of the energetics of materials under high pressures [19]. Total-energy calculations were done within the framework of the density-functional theory (DFT) and the Kohn-Sham equations were solved using the projector-augmented wave (PAW) [20, 21] method as implemented in the Vienna *ab initio* simulation package (VASP) [22]. We use a plane-wave energy cutoff of 520 eV to ensure a high precision in the calculations. It is known that DFT within the local-density approximation (LDA) or the generalized-gradient approximation (GGA) often yields incorrect results for systems with $f$ electrons with a small $f$ orbital overlap and narrow $f$ bands. The implementation of the DFT+U method has been found to improve the results in the study of cerium compounds [23]. The exchange and correlation energy was described within the GGA in the PBEsol [24] prescription. The GGA+$U$ method was used to account the strong correlation between the electrons in the Ce 4$f$ shell on the basis of Dudarev's method [25]. In this method the on-site Coulomb interaction, $U$ (Hubbard term), and the on-site exchange interaction, $J_H$, are treated together as $U_{eff} = U - J_H$. For our GGA+$U$ calculations we have chosen a value $U = 6$ eV and $J_H = 1$ eV for Ce atom. These values were chosen by comparison with the electronic structure study of $CeVO_4$ done by Da Silva *et al.* [26]. Monkhorst-Pack scheme was employed for the Brillouin-zone (BZ) integrations [27] with the grids 4x4x4, 4x4x2, and 4x4x3 for zircon, scheelite, and monazite phases, respectively. In the relaxed equilibrium configuration, the forces are less than 4 meV/Å per atom in each of the cartesian directions. Lattice-dynamics calculations of phonon modes were performed at the zone centre (Γ point) of the BZ. The calculations provided information about the frequency, symmetry and polarization vector of the vibrational modes in each structure. Highly converged results on forces are required for the calculation of dynamical matrix of lattice-dynamics calculations. We use the direct force-constant approach (or supercell method) [28]. The construction of the dynamical matrix at the Γ point of the BZ is particularly simple and involves separate calculation of the forces in which a fixed displacement from the equilibrium configuration of the atoms within the primitive unit cell is considered. Symmetry further reduces the computational efforts by reducing the number of such independent displacements in the analyzed structures. Diagonalization of the dynamical matrix provides both the frequencies of the normal modes and their polarization vectors. It allows us to identify the irreducible representations and the character of phonon modes at the Γ point.

## IV. Results

### A. High pressure XRD

At ambient conditions $CeVO_4$ crystallizes in the zircon phase (SG: $I4_1/amd$, Z = 4). The zircon structure can be described by $CeO_8$ dodecahedrons with eight similar Ce-O distances and isolated $VO_4$ tetrahedrons (see Fig.1). **Figure 2** shows the selective x-ray diffraction patterns of $CeVO_4$ at representative pressures. There were no noticeable changes in the diffraction pattern up to 4 GPa and the diffraction peaks could be indexed to zircon phase. At 5.5 GPa the appearance of many extra diffraction peaks were observed along with the weak remnant (112) diffraction peak of the zircon phase, as shown by arrow in **Fig. 2**. These changes in the diffraction patterns are indicative of a structural phase transition in $CeVO_4$ at this pressure. The new diffraction peaks of $CeVO_4$ could be assigned to the monoclinic monazite phase ($P2_1/n$). On further increase of pressure the monazite phase was found to be stable up to 12 GPa, which is the highest pressure reached in our XRD measurements. On release of pressure the monazite phase was quenched thus indicating the irreversible nature of the zircon-monazite phase transition. These results are in agreement with Raman measurements and calculations presented in the next sections.

The Rietveld refinement of all the background corrected diffraction patterns was carried out using Powdercell software [29]. The diffraction patterns up to $2\theta = 18°$ were used in order to avoid any kind of gasket interference in the structural refinements. At ambient pressure the lattice parameters for zircon-structured $CeVO_4$ are refined as $a = 7.399$ Å, $c = 6.482$ Å, which are consistent with the ICSD card No. 66033. The lattice parameters and fractional coordinates of the zircon phase of $CeVO_4$ are given in **Table I** and are also in good agreement with those reported in previous high-pressure and high-temperature XRD measurements by Range *et al.*, $a = 7.383$ Å, $c = 6.485$ Å [30].

The pressure evolution of the lattice parameters and equation of state of $CeVO_4$ in the zircon and monazite phase is shown in **Figure 3**. It can be seen that in the zircon phase the *a*-axis is more compressible than the *c*-axis as is evident from the increase of the *c/a* ratio from 0.878 at ambient pressure to 0.887 at 4 GPa. The linear compressibility of the *c*-axis ($K_c = 1.05 \times 10^{-3}$ GPa$^{-1}$) is smaller than that of the other two axes ($K_a = K_b = 3.48 \times 10^{-3}$ GPa$^{-1}$). These values are similar to those obtained in other zircon-type vanadates [6] and related oxides [31]. Due to deterioration of diffraction patterns in the monazite phase beyond 5.5 GPa, the value of β angle refined at 5.5 GPa was kept constant for further refinements. A typical Rietveld refinement fit of diffraction data at 7.2 GPa is shown in **Figure 4** and the lattice parameters and fractional coordinates are presented in **Table II**. These parameters are in close agreement with lattice parameters reported in recent XRD measurements in monazite $CePO_4$ [32]. In the monazite phase we have observed an anomalous behavior of the axial compression. From **Figure 4** it can be seen that *a*- and *b*-axis decrease with pressure while the *c*-axis increases with pressure. A similar behavior was observed in pressure-induced monazite-structured for orthophosphates [8, 9]. However, this behavior is in contrast with results in monazite-type $CePO_4$ and $LaPO_4$ [9, 32]. The different axial compression of $CeVO_4$ and $CePO_4$ could be attributed to the more distorted monazite structure of $CeVO_4$ than that of $CePO_4$. The reason is that the monazite phase of $CeVO_4$ is induced by pressure while that of $CePO_4$ occurs already at ambient pressure. Note that difference is found always when a HP monazite is compared with an ambient pressure monazite. The zircon to monazite phase transition in $CeVO_4$ exhibits a volume collapse of 8.6 % at 4 GPa which is quite large compared to those observed in orthophosphates [8, 10]. A fit of the volume vs. pressure data of the zircon phase to a second order Birch-Murnaghan equation of state ($B´_0 = 4$) gives bulk modulus of 118.9 GPa. This value is the smallest one found in orthovanadates [6], thus indicating that $CeVO_4$ is the most compressible orthovanadate to date. In the monazite phase, volume vs. pressure data fitted to third order Birch-Murnaghan equation of state gives bulk modulus $B_0$ = 142 GPa and $B'_0$ = 4.4, which is relatively close to the bulk modulus of monazite-type $CePO_4$ ($B_0$ = 122 GPa) [32]. During the fitting we have excluded the pressure-volume data point collected at 12 GPa because it deviated from the systematic behavior of the rest of the data. A possible reason of it might be the assumption that the β angle does not change under compression, which could lead to a unit-cell volume miscalculation.

The occurrence of the zircon-monazite transition instead of the zircon-scheelite transition is apparently in contradiction with results reported by Range *et al.* [30]. These authors, who used a large-volume press, found that upon compression at room temperature zircon-type $CeVO_4$ transforms to the scheelite phase. They only observed the occurrence of the monazite phase under the combined effect of pressure and temperature. To corroborate these results we have performed similar experiments using a large-volume press equipped with Bridgman anvils. The obtained results from x-ray diffraction measurements performed from samples quenched from different pressures are shown in Fig. 5 and tabulated in **Table III**. Our results show that zircon-type $CeVO_4$ transforms to the scheelite phase at 4 GPa and to the monazite phase at the same pressure but at temperatures close to 600 ºC. These results confirm the results reported by Range *et al.* [30]. The main difference between these experiments and the experiments performed using a DAC is that in the Bridgman cell a solid pressure media is used; i.e. the non-unixial stresses are more important. Therefore, our results evidence that zircon-type $CeVO_4$ undergoes a transformation to the scheelite structure at ambient temperature under non-hydrostatic compression. At high temperatures, stresses are relaxed and therefore the monazite phase can be obtained as in the DAC experiments. It is interesting to note that in the Bridgman-cell experiments, both the monazite and the scheelite phases are quenchable to ambient conditions, which confirms the non-reversibility of both transitions. The unit-cell parameters and atomic position of monazite-type and scheelite-type

CeVO$_4$ at ambient conditions are summarized in **Table IV.** For the sake of comparison the lattice parameters of the monazite [30, 33] and scheelite phases [30] reported in earlier measurements are also given.

## B. Structural calculations

The calculated total energy (E) as a function of volume of three different phases of CeVO$_4$ is shown in **Figure 6(a)**. According with the calculations, the most stable phase at ambient pressure is zircon. The calculated values of lattice parameters, fractional coordinates and bulk moduli for all the three phases are listed in **Table V**. These values are in good agreement with the experimental ones. The bulk modulus of the monazite phase is slightly underestimated by the calculations. In particular the calculated B$_0$ for monazite is 2% smaller than in low-pressure zircon, which is unusual for a HP phase. The thermodynamic phase transition between two structures occurs when the Gibbs free energy (G) is the same for both phases. In order to obtain the Gibbs free energy we use a quasi-harmonic Debye model that allows obtaining G at room temperature from calculations performed for T= 0 K [34]. **Figure 6(b)** shows the pressure dependence of the Gibbs free energy difference at T = 300 K for the scheelite and the monazite phases with respect to the zircon phase, which is taken as reference. In calculations, at non-zero temperatures all vibrational contributions are included, in contrast with 0 K calculations which neglect the vibrational contribution, zero point energy The calculations show a phase transition sequence with the formation of the monazite phase being prior to that of the scheelite phase, in good agreement with our experimental findings. Calculations also indicate that the monazite phase may transform to the scheelite phase at higher pressures. At 300 K, the calculated phase-transition pressure for the zircon-to-monazite transition is 2 GPa and the theoretical phase transition pressure for the monazite-to-scheelite transition is 3.9 GPa. Calculations done at 0 K and 600 K gave the same structural sequence, being the zircon-monazite transition pressure not affected by temperature.

However, the monazite-scheelite transition pressure increases with temperature, in agreement with the experimental results reported by Range *et al.* [30]. This positive slope can be rationalized using the Clausius-Clapeyron equation relates differences in volumes ($\Delta V$) and entropy ($\Delta S$) across a phase boundary to the slope of the phase boundary in P-T space: $dP/dT = \Delta S/\Delta V$ [35]. Since the volume per formula unit is smaller in scheelite than in monazite $CeVO_4$, the positive slope of the monazite-scheelite phase boundary indicate that the transition is also associated to an entropy decrease. To conclude this section we would like to mention that at 300 K, the calculated zircon-scheelite transition pressure agrees with the DAC experiments; however in our experiment, we have recorded data till 16 GPa and we have found no evidence for the monazite-to-scheeelite phase transition. A possible cause for the non occurrence of the monazite-scheelite transition is the presence of kinetics barriers as observed in other $ABO_4$ compounds [36]. Consequently, the atomic displacements needed to trigger the monazite-scheelite transition can only take place if the equilibrium transition pressure is sufficiently overstepped. New XRD measurements at pressures beyond 15 GPa, at different temperatures, are needed to check the possibility of the second phase transition. Note than under non hydrostatic compression the scheelite phase is found at 4 GPa, which supports the theoretical results. Apparently, non-hydrostatic stresses reduce kinetic barriers in $CeVO_4$. However, this subject is still open to new studies.

**C. Raman scattering**

**(i) Zircon structured $CeVO_4$**

At ambient conditions, $CeVO_4$ exists in the zircon structure (space group $I4_1/amd$ and point group $D_{4h}$) with two formula units per primitive cell. Group theoretical analysis predicts 12 Raman-active modes $2A_{1g} + 4B_{1g} + B_{2g} + 5E_g$ [37]. These modes can be further classified into internal ($\nu_1$-$\nu_4$) and external (translational, T, and rotational, R) modes of $VO_4$ units as follows,

$$\Gamma = A_{1g}(\nu_1, \nu_2) + B_{1g}(2T, \nu_3, \nu_4) + B_{2g}(\nu_2) + E_g(2T, R, \nu_3, \nu_4). \tag{1}$$

**Figure 7(a)** shows the Raman spectra of CeVO$_4$ in the zircon phase at different pressures up to 4.5 GPa. Eight Raman modes have been observed at ambient condition out of 12 Raman peaks predicted for CeVO$_4$ in the zircon phase. The symmetry assignment for the Raman modes has been performed in accordance with our calculations and the comparison with previous results in other vanadates and it is summarized in **Table VI**.

As can be seen from **Figure 7(a)** and **Table VI**, the intense symmetric-stretching internal mode $\nu_1(A_{1g})$, observed at 864 cm$^{-1}$ near ambient pressure, exhibits the lowest frequency for this mode when compared to other rare-earth orthovanadates [38]. This result suggests that CeVO$_4$ exhibits the weakest intra-tetrahedral V-O bonds of all orthovanadates. The two asymmetric-stretching modes $\nu_3(E_g)$ and $\nu_3(B_{1g})$ have been observed at 801 and 789 cm$^{-1}$, respectively. Apart from these phonons of CeVO$_4$, the 4 bending modes of the VO$_4$ unit could be observed in the frequency range from 260 to 500 cm$^{-1}$. However, we have observed only 3 bending modes [$\nu_2(B_{2g})$, $\nu_2(A_{1g})$, and $\nu_4(B_{1g})$] whose frequencies at ambient pressures are 262, 381, and 469 cm$^{-1}$, respectively. The asymmetric-bending mode $\nu_4(E_g)$ could not be detected. In most of the orthovanadates this mode remains undetectable [39 - 43]. Similarly, out of the five external modes two T($E_g$) modes and one T($B_{1g}$) mode have not been detected. These modes are absent probably due to their weak Raman scattering cross-sections.

**Figure 7(b)** shows the pressure dependence of Raman modes of CeVO$_4$ in the zircon phase. The symmetry assignment for the Raman modes along with their experimental and calculated frequencies, pressure coefficients, and mode Grüneisen parameters ($\gamma$) are shown in **Table VI**. A very good agreement is found between experimental and theoretical results. The frequencies of almost all Raman-active modes of the zircon phase exhibit a positive pressure coefficient. The rotational mode has the highest Grüneisen parameter, being the mode most sensitive to changes of volume. In addition, the internal bending mode of $\nu_2(B_{2g})$ symmetry shows a negative pressure coefficient. A similar soft behavior was observed for the $\nu_2(B_{2g})$ mode in other orthovanadates, e.g. YVO$_4$ [39], YbVO$_4$ [41], LuVO$_4$ [42], ScVO$_4$ [40] and also in ScPO$_4$ [10]. Our calculations also predict the softening of the T(E$_g$) mode with lowest frequency in CeVO$_4$. This softening has been also observed in other orthovanadates and it could be a characteristic behavior of zircon-type compounds [39 - 42]. It is important to note that the softening of both T(E$_g$) and $\nu_2(B_{2g})$ modes has been observed in compounds exhibiting the zircon-to-scheelite transition [39 - 44] and also in those showing the zircon-to-monazite transition [7, 10]. Therefore, it indicates that the softening of both modes is a consequence of the instability of the zircon phase. We will discuss this issue in great detail in our discussion section.

As regards the values of the pressure coefficients of the Raman-active modes of the zircon-type orthovanadates, it is worth to mention that the pressure coefficients of the internal stretching modes in CeVO$_4$ is of the order of 4.9 to 5.5 cm$^{-1}$/GPa, which is in close comparison with other orthovanadates [39 - 44]. Similarly, the pressure coefficients of the $\nu_2(B_{2g})$ bending mode is of the order of -1.2 as among orthovanadates. Finally, it is interesting to note that the pressure coefficient of the rotational R(E$_g$) mode of CeVO$_4$ has the smallest pressure coefficient of all known orthovanadates; thus indicating weak Ce - VO$_4$ bonds as compared to other orthovanadates [39 - 44].

**(ii)    Monazite structured CeVO$_4$**

Raman scattering measurements in CeVO$_4$ evidence that this compound undergoes a zircon-to-monazite phase transition above 5 GPa. Group theoretical calculations for CeVO$_4$ in monazite phase (SG: *P2$_1$/n*, point group $C_{2h}^5$) predict 36 vibrational Raman modes at the BZ centre with following symmetries $\Gamma$ = 18A$_g$ + 18 B$_g$ [45]. These modes can be further classified into internal ($\nu_1$-$\nu_4$) and external (translational, T, and rotational, R) modes of VO$_4$ units as follows,

$$\Gamma = A_g (6T, 3R, \nu_1, 2\nu_2, 3\nu_3, 3\nu_4) + B_g (6T, 3R, \nu_1, 2\nu_2, 3\nu_3, 3\nu_4) \qquad (2)$$

Raman spectra of CeVO$_4$ in the monazite phase at selected pressures are shown in **Figure 8(a)**. At 5.3 GPa we have observed the appearance of several Raman bands, both in the range between 0 and 500 cm$^{-1}$ and between 700 and 1000 cm$^{-1}$, accompanied by the broadening of many Raman modes. These changes in the Raman spectra are indicative of a structural phase transition towards the lower-symmetry monoclinic monazite phase. This result is consistent with our XRD investigations and with previous results in orthophosphates [7, 10]. Out of 36 Raman-active modes in the monazite phase, we observed only 27 modes with measurable intensity up to 12 GPa. The absence of nine Raman modes could be due to very small Raman scattering cross section and also possibly due to overlapping of many A$_g$ and B$_g$ modes due to their small splitting. Similarly, in Raman scattering measurements of several monazite-type orthophosphates 22 Raman modes were recorded out of 36 Raman active modes [46]. Above 12 GPa only the Raman-active modes of the high frequency region above 700 cm$^{-1}$ were noticeable and on further increase of pressure at 15.9 GPa all the Raman modes are completely diminished. This could have happened due to extremely weak Raman signal at this pressure or possibly due to a second phase transition to a Raman non-active phase or because of pressure-induced amorphization. Future studies will be done to clarify this point which is beyond the scope of this paper.

The mode assignment of the experimentally-observed Raman modes in CeVO$_4$ in the monazite phase was done by comparing the experimental and calculated values of frequencies and pressure coefficients and is shown in **Table VII** together with mode Grüneisen parameters ($\gamma$), calculated using the bulk modulus of the monazite phase, B$_0$ = 131.5 GPa, obtained from XRD measurements. The assignment of modes can be done broadly in two regions. The internal modes of vibration of VO$_4$ tetrahedra covered the frequency range of 290-950 cm$^{-1}$ while the external modes of VO$_4$ unit span the frequency range of 50-270 cm$^{-1}$. This assignment is consistent in comparison with orthophosphates [46]. In general, the calculated Raman frequencies and their pressure coefficients are in good agreement with our experimental ones. The only mode whose assignment is doubtful is the soft mode observed at 89.6 cm$^{-1}$ at ambient pressure. This mode shows a frequency close to that of the calculated $A_g$ mode at 92.5 cm$^{-1}$ but its negative pressure coefficient is more similar to that of the calculated $B_g$ mode at 95.3 cm$^{-1}$.

**Figure 8(b)** shows the pressure dependence of the Raman-active modes of CeVO$_4$ in the monazite phase. All the experimental Raman-active modes show a positive pressure coefficient except the two Raman modes at 89.6 cm$^{-1}$ and 140.4 cm$^{-1}$ which is in good agreement with our calculations that predict three soft modes. The pressure-evolution observed for the Raman modes is similar to that reported for monazite CePO$_4$ [32]. Only the soft mode of $B_g$ symmetry calculated at 66.2 cm$^{-1}$ at ambient pressure was not measured. The softening of the same Raman modes was previously observed in TbVO$_4$ [8]. On release of pressure from 15.9 GPa to ambient pressure the monazite phase was recovered, thus showing the irreversible nature of the zircon-to-monazite phase transition in good agreement with our XRD results. It must be noted that a similar behavior was observed for TbPO$_4$ [7].

**(iii) Scheelite structured CeVO$_4$**

Group theoretical calculations for CeVO$_4$ in the scheelite phase (SG: $I4_1/a$, point group $C_{4h}^6$) predict 13 vibrational Raman modes at the BZ centre with following symmetries $\Gamma = 3A_g + 5B_g + 5E_g$ [47]. As already commented, our theoretical calculations predict the possibility of the scheelite phase for CeVO$_4$ which has been observed in Bridgman-cell experiments. **Table VIII** shows the comparison of theoretically calculated and experimental Raman modes of CeVO$_4$ in the scheelite phase along with pressure coefficients and with mode Grüneisen parameters ($\gamma$), to evaluate the mode Grüneisen parameters ($\gamma$) the calculated bulk modulus $B_0 = 138.65$ GPa is used. The calculated Raman frequencies match quite well with the experimental ones at ambient conditions. The pressure coefficients are similar to that observed in the scheelite phase in other vanadates [39 – 43]. In **Table VIII**, it can be seen that the strongest mode of scheelite, the symmetric stretching $A_g$ mode, is observed at 826.7 cm$^{-1}$, which is consistent with other scheelite orthovanadates [39, 40]. This implies a collapse of the frequency of the strongest mode in comparison with the zircon structure, a typical feature of the zircon-scheelite transition. Finally, the frequencies of other Raman modes are also in quite agreement with other scheelite orthovanadates [39, 40].

V.  **Discussion**

Zircon-structured orthovanadates under hydrostatic compression were observed to undergo the zircon-to-scheelite and then the scheelite-to-fergusonite phase transitions [39 - 42]. In all cases, the scheelite phase was found to be quenchable on release of pressure. The zircon-to-scheelite phase transition is a rather sluggish first-order and reconstructive transition in nature [48] while the scheelite-to-fergusonite is a ferroelastic second-order and displacive transition [49]. In our case, the zircon to scheelite transition involves a volume collapse of 11%, which is consistent with the first-order character of the transformation. On the other hand, the zircon-to-monazite phase transition is a rather sudden transition with a volume collapse of 8%, with an increase of the $Ce^{3+}$ coordination, and without coexistence of phases over a wide pressure range (typical of the zircon-scheelite transition). All these facts indicate the first-order and reconstructive nature of this phase transition. The structural relationships between monazite and zircon were elucidated by Ni *et al.* [5]. These relationships can be used to understand the reconstructive nature of the zircon-to-monazite phase transition with the help of **Fig. 1**. In monazite, $CeO_9$ polyhedra share edges and corners with isolated $VO_4$ tetrahedra whereas $CeO_8$ dodecahedra in zircon. The edge sharing $VO_4$ and $CeO_8$ polyhedral chains along the [001] direction of the zircon structure appears to be like polyhedral chains in monazite, although the chains are twisted to accommodate the ninth atom. However, due to the extra ninth oxygen atom packing efficiency in monazite phase is better compared to zircon phase. This result supports the observed higher bulk modulus value for monazite CeVO$_4$. This new atomic arrangement in the monazite phase can be realized by considering a slight shift of the [001] planes and a rotation of VO$_4$ polyhedra in the *a-b* plane. The fact that this transition (as well as the zircon-scheelite transition) is not reversible is a notable aspect. We think that a large kinetic barrier is the cause of non reversibility, which is consistent with the reconstructive mechanism of this transition and its first-order character. Another important issue to remark is that the presence of the AO$_9$ polyhedra (a cationic pentagonal interpenetrating tetrahedral polyhedral) allows monazite to accommodate chemically diverse cations. The irregular coordination around the A cation does not place severe symmetry, size, or charge constrains and allows large domains of chemical composition. Therefore, we predict that in addition to CeVO$_4$, other zircon-structured vanadates with large A cations could also take the monazite structure under pressure, becoming isostructural to LaCrO$_4$.

An important issue related with the fact that under compression the zircon structure becomes mechanically unstable is the softening of external translational T(E$_g$) and internal $\nu_2$(B$_{2g}$) bending modes. In zircon structured compounds, the softening of the T(E$_g$) mode is related with a softening of the C$_{44}$ elastic constant. This fact is caused by monoclinic distortions in the *a-c* or *b-c* planes [50]. Similarly, the softening of $\nu_2$(B$_{2g}$) mode is related to a softening of the C$_{66}$ elastic constant, which is caused by orthorhombic distortions in the basal plane [50]. Low-temperature studies of zircon-structured DyVO$_4$, DyAsO$_4$ and TbVO$_4$ indicate a similar phonon softening which can be attributed to the distortion of the zircon structure along [110] or [100] direction [51, 52]. By analogy, the pressure-induced contraction would lead to similar distortions in the zircon structure making this structure unstable.

In order to contribute to the systematic understanding of structural properties of zircon-type oxides, here we will make the attempt to understand the factors which govern the phase stability at ambient pressure and the sequence of pressure-induced structural phase transitions in orthovanadates and orthophosphates. As commented earlier, the phase stability of zircon-type orthovanadates and orthophosphates and their phase transitions seem to greatly depend upon the ionic radii. This is consistent with the recent updated version of Bastide's diagram for $ABX_4$ compounds elaborated and discussed by D. Errandonea and F.J. Manjón [53]. In Bastide´s diagram, the phase stability and phase transitions of $ABO_4$ orthovanadates and orthophosphates can be understood by considering the role played by cationic radii, $r_A$ and $r_B$, with respect to anion radius, $r_O$ [54]. In this diagram, those compounds which have $r_A/r_O$ and $r_B/r_O$ cation-to-anion ratios well inside the stability region of the zircon structure are observed to undergo the zircon-to-scheelite phase transition and follow the traditional north-east (NE) rule in the Bastide's diagram [54]. Such is the case of many zircon-structured orthovanadates, like $YVO_4$ [39], $ScVO_4$ [40], $YbVO_4$ [41], $LuVO_4$ [42], and also in $ScPO_4$ [10]. On the other hand, those compounds whose $r_A/r_O$ and $r_B/r_O$ cation-to-anion ratios fall near border of the stability region of the zircon and monazite structures could crystallize in both phases, like $CeVO_4$, $TbPO_4$ [7, 8], and $YPO_4$ [9]. In this case, the compounds crystallizing in the zircon phase, like $CeVO_4$, are prone to undergo the zircon-to-monazite phase transition. Few orthophosphates, like $YbPO_4$ and $LuPO_4$ [11], are exception to this rule and deserve further investigation. Finally, the last case would correspond to compounds whose $r_A/r_O$ and $r_B/r_O$ cation-to-anion ratios fall well inside the stability region of the monazite phase. This is the case for most orthophosphates which already crystallize in the monazite phase. To conclude we would like to add that the crystal chemistry arguments here used to discuss the structural behavior of phosphates and vanadates has been recently satisfactory use to describe arsenates and chromates [55].

Finally, we would like to comment on additional consequences of the observed pressure-driven structural changes. $CeVO_4$ is a large band-gap material with a band-gap energy of 3.1-4.2 [56, 57]. It is usually assumed that the electronic structure near the Fermi level is dominated by V $3d$ and O $2p$ states. However, recently a band-gap of 1.8 eV was measured [58] being attributed to the presence of localized $4f$ levels of Ce between the valence and conduction bands. Clearly there is a contradiction between both pictures of the electronic band structure which could be solved by HP optical-absorption studies like those already preformed in $PbWO_4$ [59]. It is known that pressure affects differently localized and delocalized electronic states. In particular, if $4f$ states are present near the Fermi level, we expect $E_g$ to be considerable reduced by pressure within the stability range of the zircon structure. In addition, both the zircon-scheelite and zircon-monazite transition will probably cause a large collapse of the band-gap as a consequence of the atomic rearrangement after the phase transitions. Finally, another issue interesting to explore in the future are the effects caused by HP $f$ electron delocalization of lanthanides [60], which, among other things, should modify the magnetic properties of compounds like $CeVO_4$.

**VI. Conclusion**

X-ray diffraction, Raman scattering, and theoretical studies of $CeVO_4$ up to 16 GPa suggests that the low-pressure zircon phase undergo an irreversible zircon-to-monazite phase transition at 5.3 GPa. XRD and Raman signals weaken considerably beyond 12 and 16 GPa, respectively, evading any further measurements. The symmetries of the Raman modes in the zircon and monazite phases of $CeVO_4$ have been assigned in accordance with our lattice dynamics calculations. In general, good agreement is found for the zircon to monazite phase transition by our experimental and theoretical data. Our calculations predict the monazite to scheelite phase transition which could not be detected experimentally under hydrostatic conditions. However, the scheelite structure is observed upon compression under non-hydrostatic conditions. The softening of external translational T($E_g$) and internal $\nu_2$($B_{2g}$) bending modes can be attributed to distortion in the zircon structure along [110] or [100] direction associated with the phase transition. We have also discussed the dependence of the phase stability of the zircon phase and its pressure-induced phase transitions either to the scheelite or monazite phases in orthovanadates and orthophosphates on their $r_A/r_O$ and $r_B/r_O$ cation-to-anion ratios. Finally, the effects of non hydrostaticity in the structural sequence are discussed. We found that different compression methods induced a transition to different structures.

**Acknowledgments**

Research financed by the Spanish MCINN under Grants No. MAT2010-21270-C04-01/02/04, CSD2007-00045, and No. CTQ2009-14596-C02-01. Some of the authors are members of the MALTA Consolider Team. S. L-M acknowledges the financial support of the CONACyT México under a postdoctoral fellowship. Supercomputer time has been provided by the Red Española de Supercomputación (RES) and the MALTA cluster.

**Table I:** Lattice parameters and fractional coordinates of CeVO$_4$ in zircon phase at ambient conditions. Space group *I4$_1$/amd*, Z= 4, *a* = 7.399 Å, *c* = 6.482 Å and residuals R$_p$ = 11.66 R$_{wp}$ = 11.24.

| Atoms | Sites | x | y | z |
|---|---|---|---|---|
| Ce | 4a | 0 | 0.7500 | 0.1250 |
| V | 4b | 0 | 0.2500 | 0.3750 |
| O | 16h | 0 | 0.4264 | 0.2134 |

**Table II:** Lattice parameters and fractional coordinates of CeVO$_4$ in monazite phase at 5.5 GPa. Space group *P2$_1$/n*, Z= 4, *a* = 6.980 Å, *b* = 7.079 Å, *c* = 6.550 Å and β = 105.3° and residuals R$_p$ = 32.46  R$_{wp}$ = 37.99.

| Atoms | Sites | x | y | z |
|---|---|---|---|---|
| Ce | 4e | 0.2818 | 0.1591 | 0.1000 |
| V | 4e | 0.3047 | 0.1635 | 0.6124 |
| O$_1$ | 4e | 0.2508 | 0.0055 | 0.4458 |
| O$_2$ | 4e | 0.3811 | 0.3320 | 0.4982 |
| O$_3$ | 4e | 0.4745 | 0.1054 | 0.8042 |
| O$_4$ | 4e | 0.1268 | 0.2164 | 0.7108 |

**Table III:** Results obtained from Bridgman-cell experiments.

| Experiment number | Temperature (°C) | Pressure (GPa) | Recovered Sample Phase |
|---|---|---|---|
| Run 1 | 25 | 2 | Zircon |
| Run 2 | 600 | 2 | Zircon |
| Run 3 | 25 | 4 | Scheelite |
| Run 4 | 600 | 4 | Monazite |
| Run 5 | 25 | 6 | Scheelite |
| Run 6 | 600 | 6 | Scheelite |

**Table IV:** Lattice parameters of $CeVO_4$ in scheelite phase and monazite phase obtained from Bridgman-cell compression with temperature variation. For comparison purpose lattice parameters reported by Range *et. al.* [30] and Yoshimura *et. al.* [33] are also given.

| Monazite phase ($P2_1/n$) | | | Scheelite phase ($I4_1/a$) | |
|---|---|---|---|---|
| This work | Ref. 30 | Ref. 33 | This work | Ref. 30 |
| $a$ = 7.001 Å | $a$ = 7.003 Å | $a$ = 6.98 Å | $a$ = 5.163 Å | $a$ = 5.165 Å |
| $b$ = 7.221 Å | $b$ = 7.227 Å | $b$ = 7.22 Å | $c$ = 11.849 Å | $c$ = 11.848 Å |
| $c$ = 6.703 Å | $c$ = 6.685 Å | $c$ = 6.76 Å | | |
| $\beta$ = 105.07 | $\beta$ = 105.13 | $\beta$ = 105.02 | | |

**Table V:** Calculated lattice parameters, bulk modulus, and fractional coordinates of CeVO$_4$ in zircon phase at ambient conditions. Space group $I4_1/amd$, Z= 4, $a = 7.423$ Å, $c = 6.461$ Å, and the bulk modulus B$_0$ = 115.24 GPa, B´$_0$ = 4.83

| Atoms | Sites | x | y | z |
|---|---|---|---|---|
| Ce | 4a | 0 | 0.7500 | 0.1250 |
| V | 4b | 0 | 0.2500 | 0.3750 |
| O | 16h | 0 | 0.4289 | 0.2053 |

Calculated lattice parameters, bulk modulus and fractional coordinates of CeVO$_4$ in the monazite phase at 5.77 GPa. Space group $P2_1/n$, Z= 4, $a = 6.869$ Å, $b = 7.119$ Å, $c = 6.56$ Å, $\beta = 104.88°$ and the bulk modulus B$_0$ = 109.63 GPa, B´$_0$ = 3.24.

| Atoms | Sites | x | y | z |
|---|---|---|---|---|
| Ce | 4e | 0.2865 | 0.1581 | 0.1086 |
| V | 4e | 0.3002 | 0.1681 | 0.6184 |
| O$_1$ | 4e | 0.2393 | 0.9971 | 0.4264 |
| O$_2$ | 4e | 0.3883 | 0.3489 | 0.4949 |
| O$_3$ | 4e | 0.4902 | 0.1113 | 0.8327 |
| O$_4$ | 4e | 0.1158 | 0.2269 | 0.7386 |

Calculated lattice parameters, bulk modulus, and fractional coordinates of CeVO$_4$ in the scheelite phase at ambient conditions. Space group $I4_1/a$, Z= 4, $a = 5.165$ Å and $c = 11.795$ Å, and the bulk modulus B$_0$ = 138.65 GPa, B´$_0$ = 3.84.

| Atoms | Sites | x | y | z |
|---|---|---|---|---|
| Ce | 4b | 0 | 0.2500 | 0.6250 |
| V | 4a | 0 | 0.2500 | 0.1250 |
| O | 16f | 0.2500 | 0.1182 | 0.0463 |

**Table VI:** *Ab initio* calculated and experimental frequencies at ambient conditions (0.5 GPa), pressure coefficients, and mode Grüneisen parameters ($\gamma$) of the CeVO$_4$ in zircon phase.

| Raman Mode Symmetry | $\omega_0^a$ [cm$^{-1}$] | $d\omega/dP^a$ [cm$^{-1}$/GPa] | $\gamma^a$ | $\omega_0^b$ [cm$^{-1}$] | $d\omega/dP^b$ [cm$^{-1}$/GPa] | $\gamma^b$ |
|---|---|---|---|---|---|---|
| T(E$_g$)    | 110.5 | -0.52 | -0.63 | ----- | ----- | ----- |
| T(B$_{1g}$) | 123.4 | 0.59  | 0.54  | 124.4 | 0.56  | 0.54  |
| T(E$_g$)    | 153.0 | 0.66  | 0.57  | ----- | ----- | ----- |
| R(E$_g$)    | 232.5 | 4.92  | 2.61  | 234.1 | 3.52  | 1.79  |
| T(B$_{1g}$) | 236.7 | 2.92  | 1.58  | ----- | ----- | ----- |
| $\nu_2$(B$_{2g}$) | 254.1 | -1.18 | -0.97 | 261.9 | -1.25 | -0.57 |
| $\nu_4$(E$_g$)    | 354.6 | 0.42  | 0.20  | 381.1 | 0.12  | 0.04  |
| $\nu_2$(A$_{1g}$) | 368.7 | 1.46  | 0.52  | ----- | ----- | ----- |
| $\nu_4$(B$_{1g}$) | 451.4 | 2.27  | 0.65  | 468.9 | 2.26  | 0.57  |
| $\nu_3$(B$_{1g}$) | 828.3 | 5.08  | 0.80  | 789.1 | 5.26  | 0.79  |
| $\nu_3$(E$_g$)    | 828.7 | 5.41  | 0.85  | 801.3 | 4.95  | 0.73  |
| $\nu_1$(A$_{1g}$) | 872.3 | 4.37  | 0.66  | 864.3 | 5.52  | 0.76  |

[a] Theoretical calculations
[b] Experimental data

**Table VII:** Theoretical and experimental frequencies, pressure coefficients, and mode Grüneisen parameters of monazite CeVO$_4$. Grüneisen parameters, $\gamma = (B_0/\omega_0) \times d\omega/dP$, calculated assuming $B_0 = 142$ GPa.

| Raman Mode Symmetry | $\omega^a$ [cm$^{-1}$] | $d\omega/dP^a$ [cm$^{-1}$/GPa] | $\gamma^a$ | $\omega^b$ [cm$^{-1}$] | $d\omega/dP^b$ [cm$^{-1}$/GPa] | $\gamma^b$ |
|---|---|---|---|---|---|---|
| B$_g$ | 66.2 | -0.58 | -1.11 | ----- | ----- | ----- |
| A$_g$ | 73.5 | 0.51 | 0.82 | 72.1 | 0.35 | 0.69 |
| A$_g$ | 92.5 | 0.27 | 0.35 | 89.6 | -0.31 | -0.49 |
| B$_g$ | 95.3 | -1.20 | -1.61 | ----- | ----- | ----- |
| A$_g$ | 107.3 | 0.01 | 0.01 | 104.2 | 0.13 | 0.18 |
| B$_g$ | 121.6 | 1.35 | 1.30 | 126.3 | 0.55 | 0.62 |
| B$_g$ | 126.4 | 1.28 | 1.18 | ----- | ----- | ----- |
| A$_g$ | 136.6 | -0.30 | -0.27 | 140.4 | -0.69 | -0.69 |
| A$_g$ | 147.4 | 1.67 | 1.32 | 146.5 | 1.26 | 1.22 |
| A$_g$ | 159.9 | 2.34 | 1.69 | 161.8 | 0.95 | 0.83 |
| B$_g$ | 160.7 | 2.35 | 1.68 | ----- | ----- | ----- |
| B$_g$ | 189.6 | 2.89 | 1.75 | 193.4 | 1.97 | 1.45 |
| A$_g$ | 193.8 | 2.81 | 1.66 | ----- | ----- | ----- |
| B$_g$ | 211.9 | 3.24 | 1.76 | 208.3 | 1.68 | 1.15 |
| B$_g$ | 233.0 | 2.90 | 1.44 | ----- | ----- | ----- |
| A$_g$ | 236.6 | 3.53 | 1.72 | 243.8 | 2.16 | 1.26 |
| B$_g$ | 246.2 | 3.94 | 1.83 | ----- | ----- | ----- |
| A$_g$ | 259.7 | 3.29 | 1.46 | 258.2 | 2.56 | 1.41 |
| B$_g$ | 296.5 | 0.96 | 0.39 | 310.1 | 1.58 | 0.72 |
| A$_g$ | 316.5 | 0.69 | 0.27 | 326.5 | 0.37 | 0.16 |
| B$_g$ | 317.9 | 1.98 | 0.74 | ----- | ----- | ----- |
| A$_g$ | 337.4 | 2.78 | 0.97 | 334.8 | 1.57 | 0.67 |
| A$_g$ | 362.7 | 3.37 | 1.09 | 350.9 | 2.67 | 1.08 |
| A$_g$ | 382.7 | 3.67 | 1.12 | 375.8 | 2.60 | 0.98 |
| B$_g$ | 390.1 | 2.50 | 0.76 | 404.4 | 2.40 | 0.84 |
| B$_g$ | 414.1 | 2.42 | 0.70 | 425.6 | 1.72 | 0.57 |
| A$_g$ | 429.8 | 2.63 | 0.72 | 441.7 | 2.02 | 0.65 |
| B$_g$ | 433.9 | 3.81 | 1.03 | 465.0 | 3.68 | 1.12 |
| A$_g$ | 801.3 | 3.99 | 0.59 | 771.2 | 3.04 | 0.56 |
| B$_g$ | 817.2 | 3.93 | 0.57 | 784.5 | 2.90 | 0.53 |
| A$_g$ | 825.3 | 4.41 | 0.64 | 794.1 | 1.86 | 0.33 |
| A$_g$ | 853.3 | 2.90 | 0.41 | 817.5 | 1.59 | 0.28 |
| B$_g$ | 862.1 | 4.44 | 0.61 | 825.2 | 1.72 | 0.29 |
| B$_g$ | 872.8 | 3.94 | 0.54 | 854.4 | 3.17 | 0.53 |
| A$_g$ | 873.8 | 3.03 | 0.41 | 860.2 | 3.15 | 0.52 |
| B$_g$ | 903.7 | 2.29 | 0.31 | ----- | ----- | ----- |

[a] Theoretical calculations, [b] Experimental data

**Table VIII:** *Ab initio* calculated and experimental frequencies at ambient pressure for the scheelite phase.

| Raman Mode Symmetry | $\omega_0^a$ [cm$^{-1}$] | $d\omega/dP^a$ [cm$^{-1}$/GPa] | $\gamma^a$ | $\omega_0^b$ [cm$^{-1}$] |
|---|---|---|---|---|
| T(E$_g$) | 106.9 | -0.22 | -0.35 | ----- |
| T(B$_g$) | 132.0 | -0.77 | -1.04 | ----- |
| T(E$_g$) | 165.4 | 2.65 | 2.42 | ------ |
| T(B$_g$) | 176.9 | 2.80 | 2.37 | ------ |
| R(A$_g$) | 206.4 | 0.52 | 0.42 | 230.2 |
| R(E$_g$) | 273.1 | 2.57 | 1.48 | 313.5 |
| $\nu_2$(A$_g$) | 331.5 | 2.67 | 1.28 | 346.6 |
| $\nu_2$(B$_g$) | 349.1 | 1.30 | 0.61 | 367.7 |
| $\nu_4$(B$_g$) | 370.9 | 2.96 | 1.29 | 403.4 |
| $\nu_4$(E$_g$) | 393.6 | 2.70 | 1.11 | 427.8 |
| $\nu_3$(E$_g$) | 772.6 | 3.61 | 0.76 | 742.3 |
| $\nu_3$(B$_g$) | 817.1 | 3.09 | 0.62 | 799.8 |
| $\nu_1$(A$_g$) | 840.26 | 2.71 | 0.54 | 826.7 |

[a] Theoretical calculations, [b] Experimental data

**Figure Captions**

**Figure 1:** (a) Schematic view of the zircon structure. (b) Schematic view of the monazite structure. (c) Schematic view of the scheelite structure. Black solid spheres correspond to Ce-atoms, dark-grey solid spheres correspond to O-atoms, and white solid spheres correspond to V-atoms. The different polyhedra are illustrated in all figures (left $VO_4$ octahedra, right $CeO_8$ or $CeO_9$ polyhedra).

**Figure 2**: Evolution of the x-ray diffraction patterns of $CeVO_4$ as a function of pressure. Asterisks represent the diffraction peaks due to gasket. Arrow indicates the remnant (112) diffraction peak of the zircon phase.

**Figure 3**: Pressure dependence of the lattice parameters and volume. Filled circles (squares) correspond to the zircon (monazite) phase of $CeVO_4$. The solid lines in the lattice-parameter plots are linear fits of the data and the solid lines in the volume data correspond to the third order Birch-Murnaghan EOS.

**Figure 4**: Observed and calculated x-ray diffraction patterns for the monazite *($P2_1/n$, $Z = 4$)* phase of $CeVO_4$ at 7.3 GPa. Bars indicate the expected positions of diffraction peaks.

**Figure 5:** XRD patters collected from different phases of $CeVO_4$ recovered after experiments in the Bridgman cell.

**Figure 6**: (a) Variation of total energy, at T= 0 K, as a function of volume for zircon, scheelite and monazite-type $CeVO_4$. Filled circles, squares, and triangles correspond to zircon, monazite, and scheelite phases, respectively. (b) Plot of the free energy, at T = 300 K, versus pressure for the scheelite and the monazite phases. The free energy of zircon phase has been taken as a reference.

**Figure 7**: (a) Raman spectra of CeVO$_4$ in the zircon phase between 0.5 GPa and 4.5 GPa. (b) Experimental pressure dependence of the Raman-mode frequencies in zircon-type CeVO$_4$. The solid lines are the calculated modes. The dashed lines represent Raman modes not observed in the experiments.

**Figure 8**: (a) Raman spectra of the monazite phase of CeVO$_4$ at pressures between 5.3 GPa and 15.9 GPa. (b) Experimental pressure dependence of the Raman-mode frequencies in monazite phase CeVO$_4$. Filled and empty squares correspond to Raman modes in monazite CeVO$_4$ on upstroke and downstroke, respectively. The solid lines are the calculated modes. The dashed lines represent Raman modes not observed in the experiments.

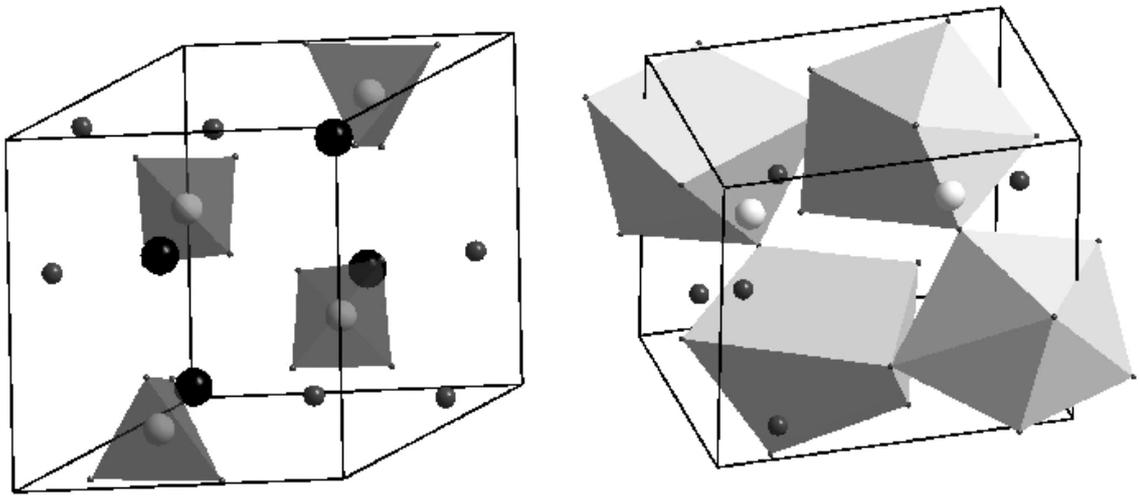

Figure 1(a)

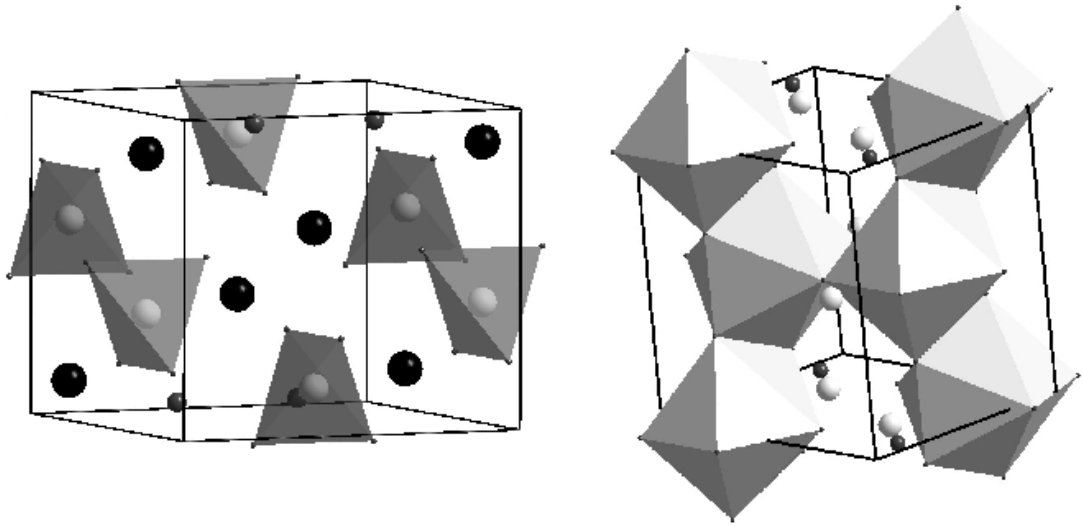

Figure 1(b)

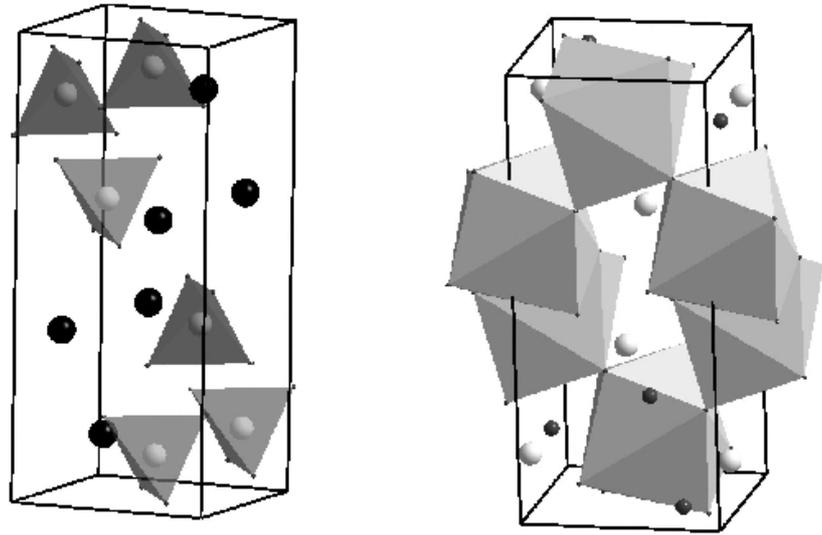

Figure 1(c)

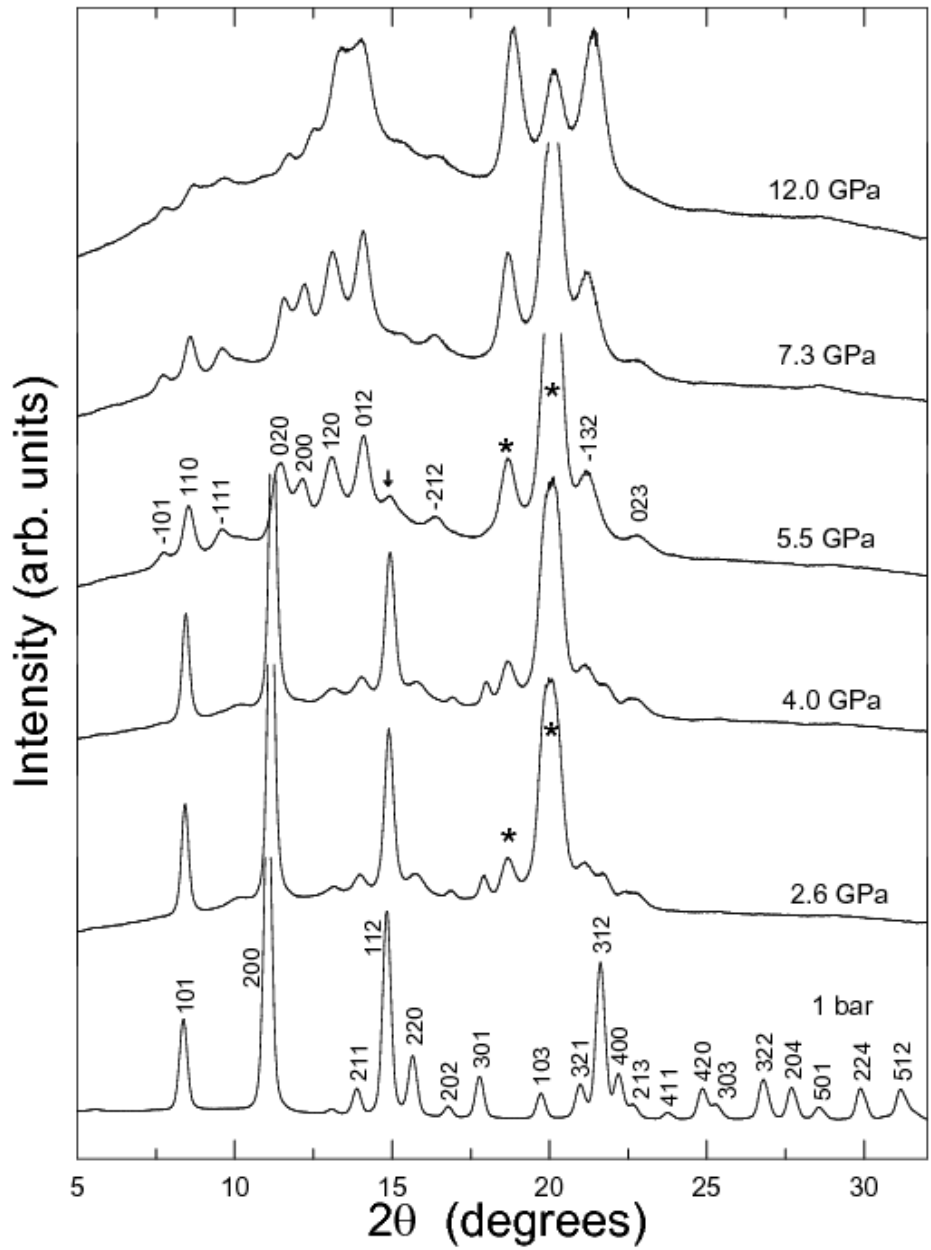

Figure 2

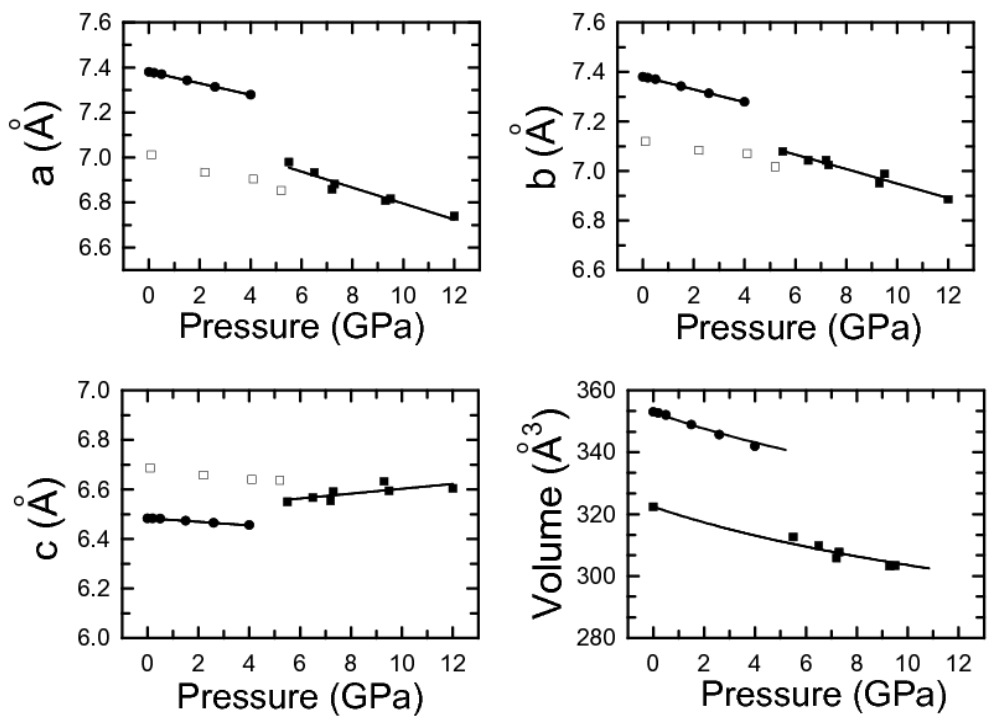

Figure 3

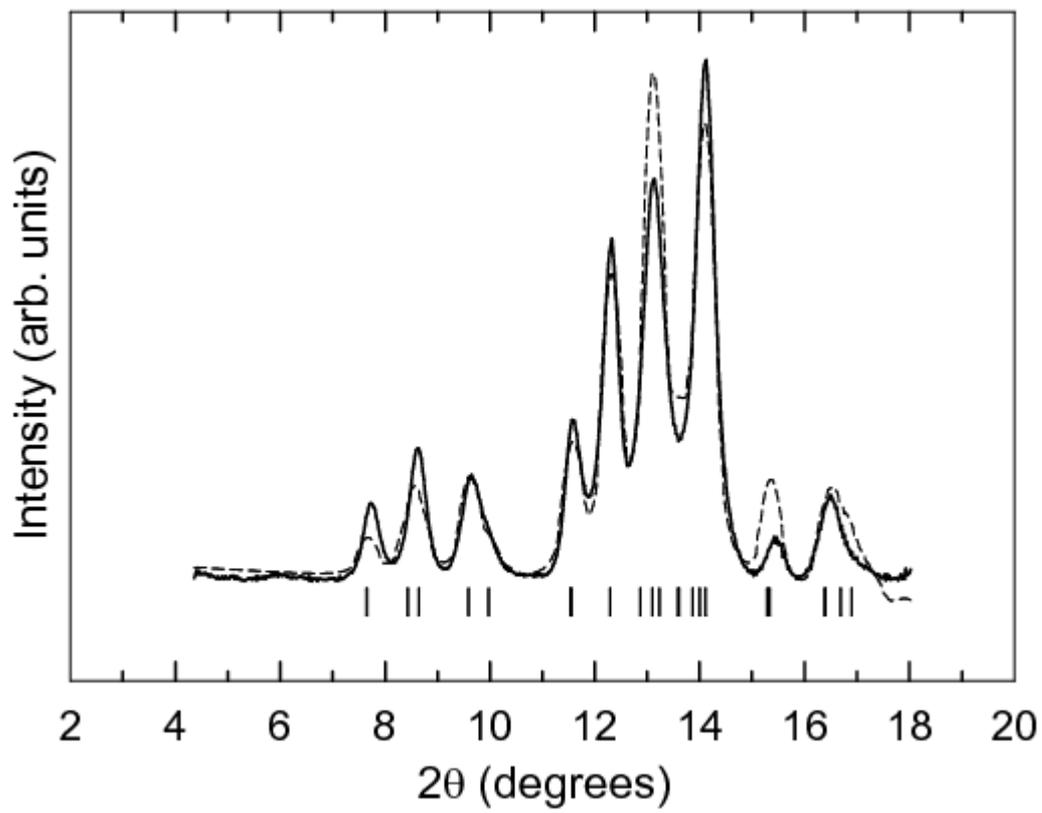

Figure 4

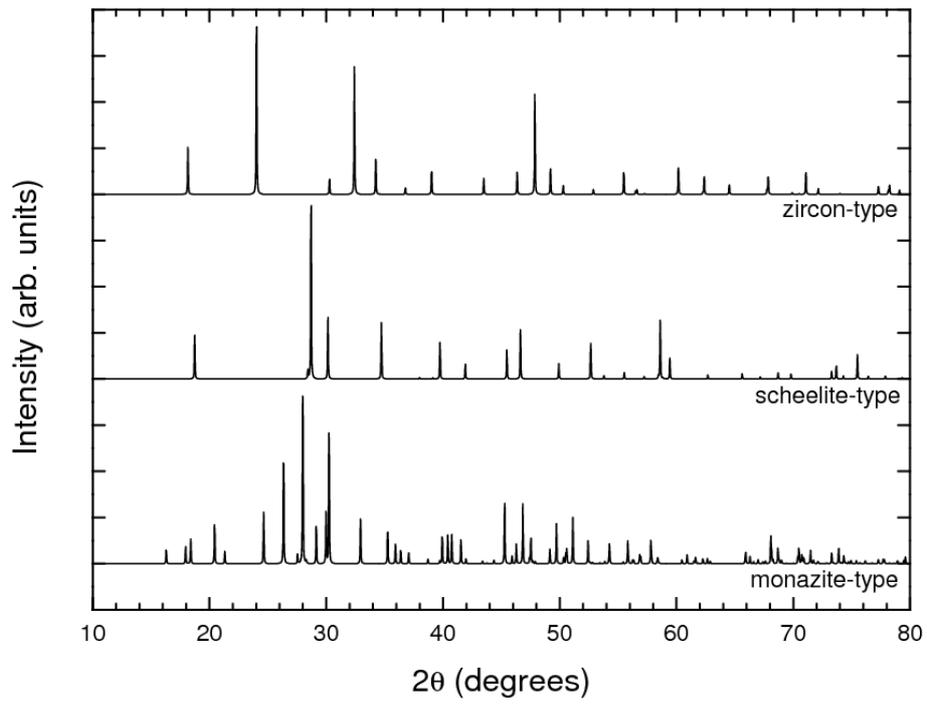

Figure 5

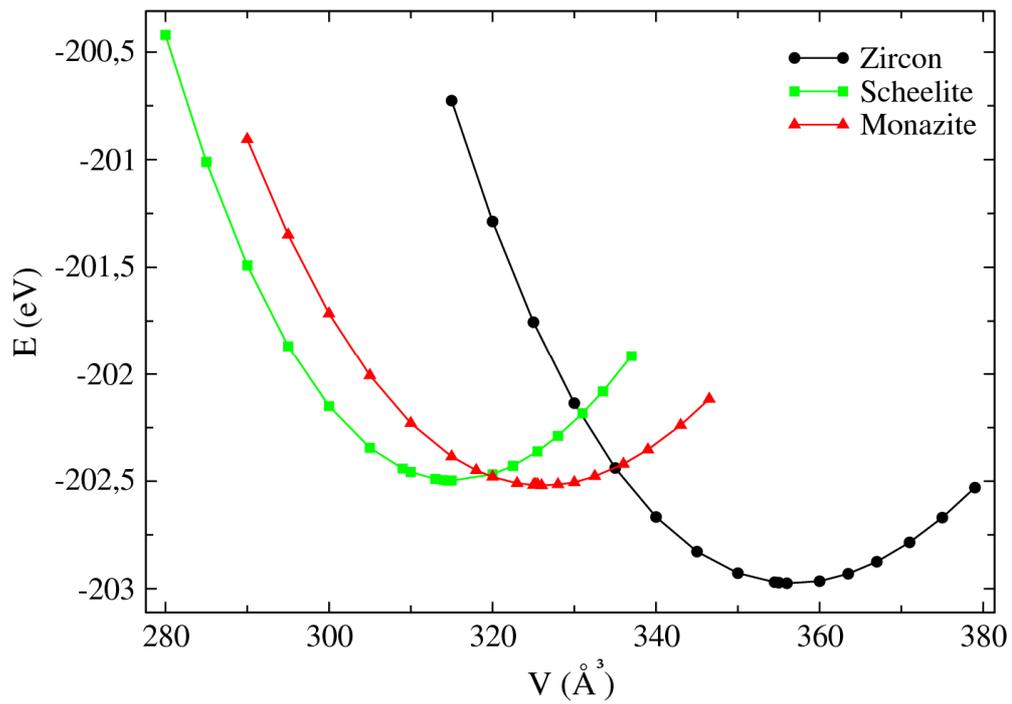

Figure 6(a)

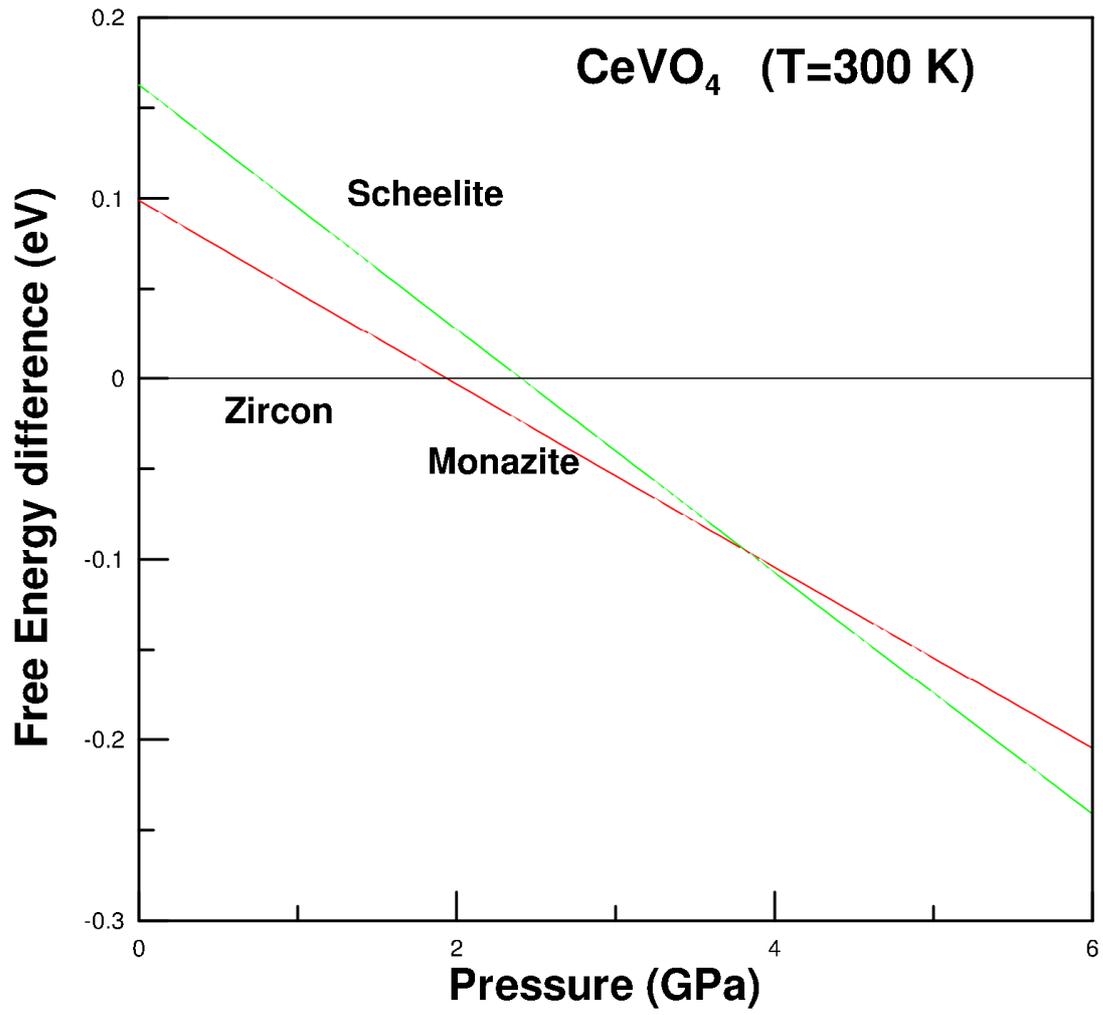

Figure 6(b)

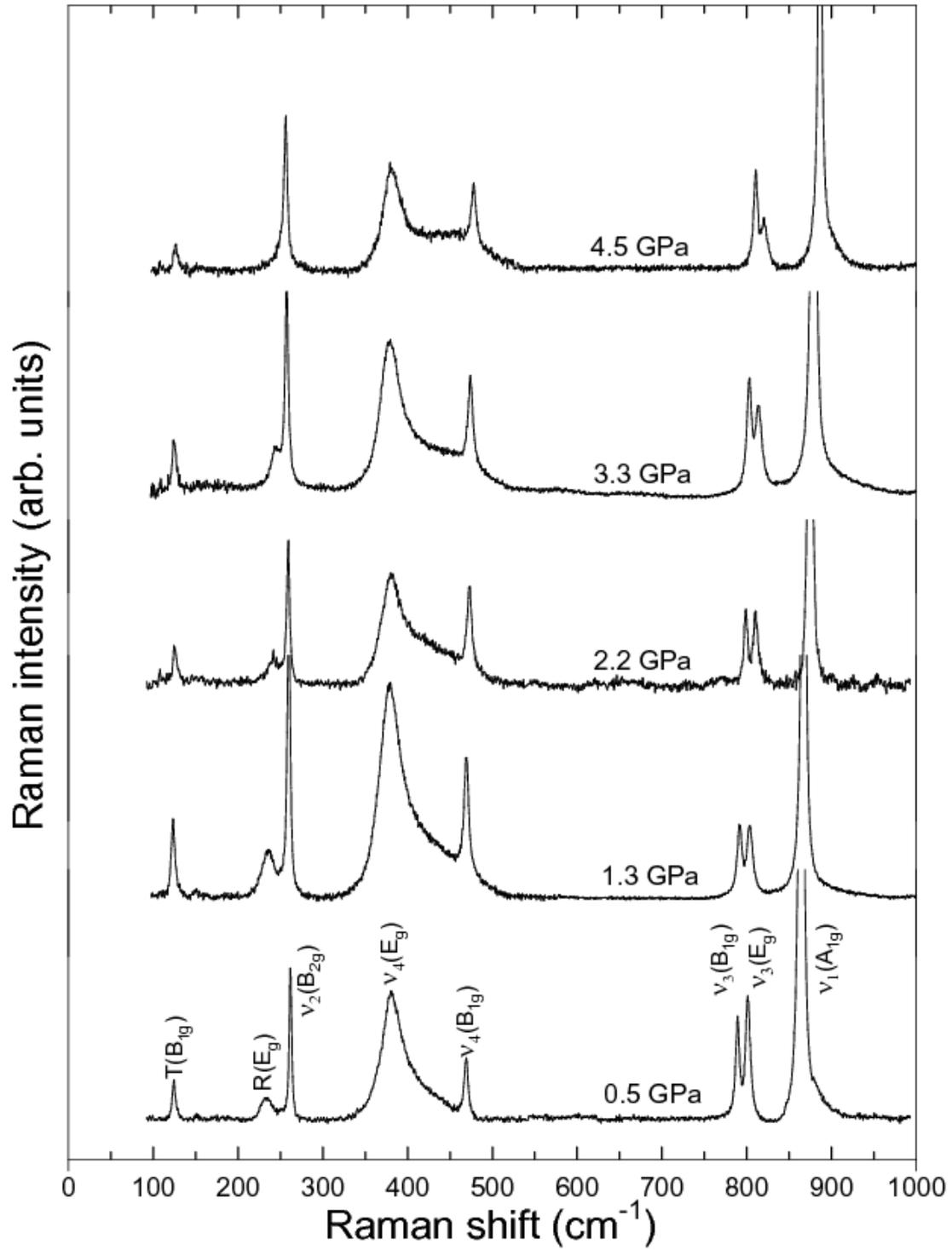

Figure 7(a)

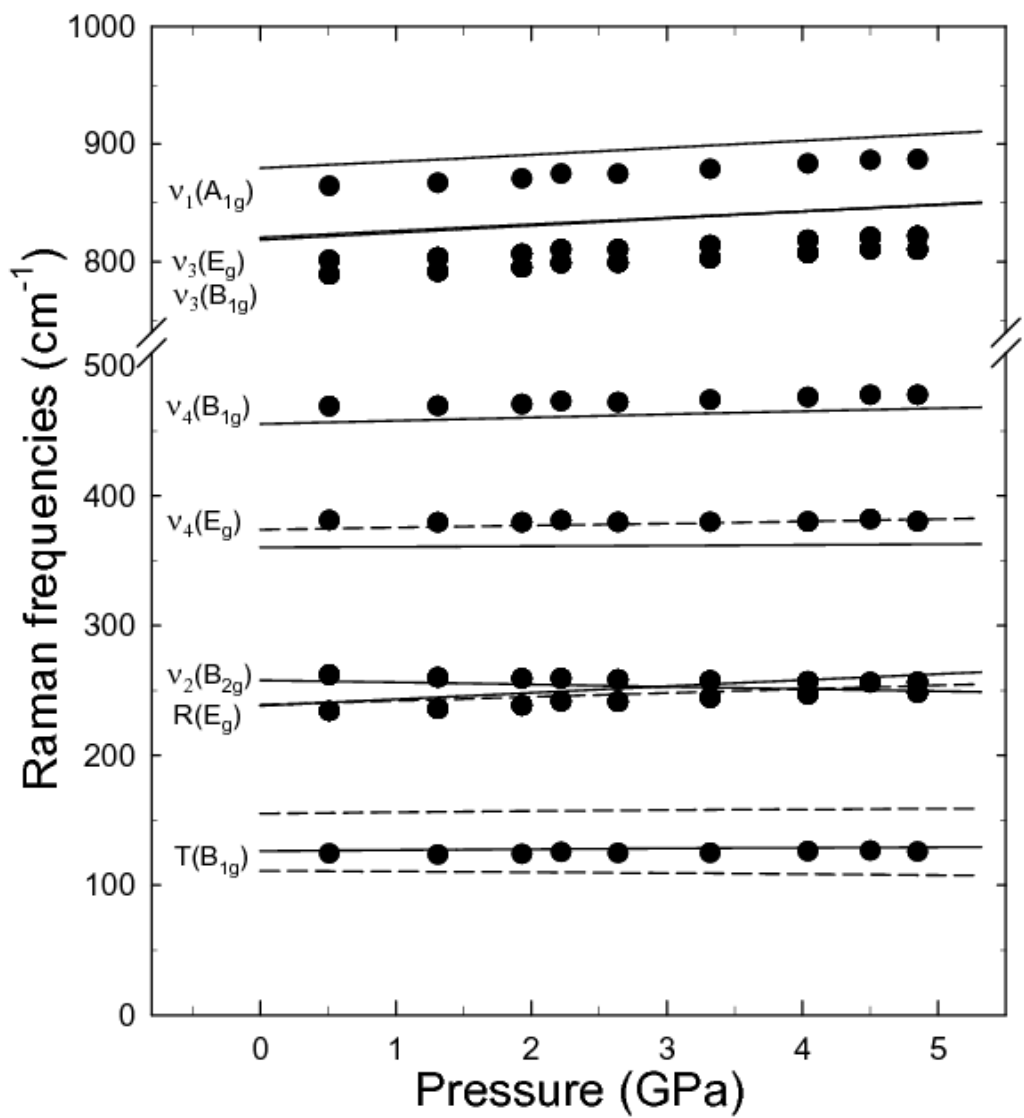

Figure 7(b)

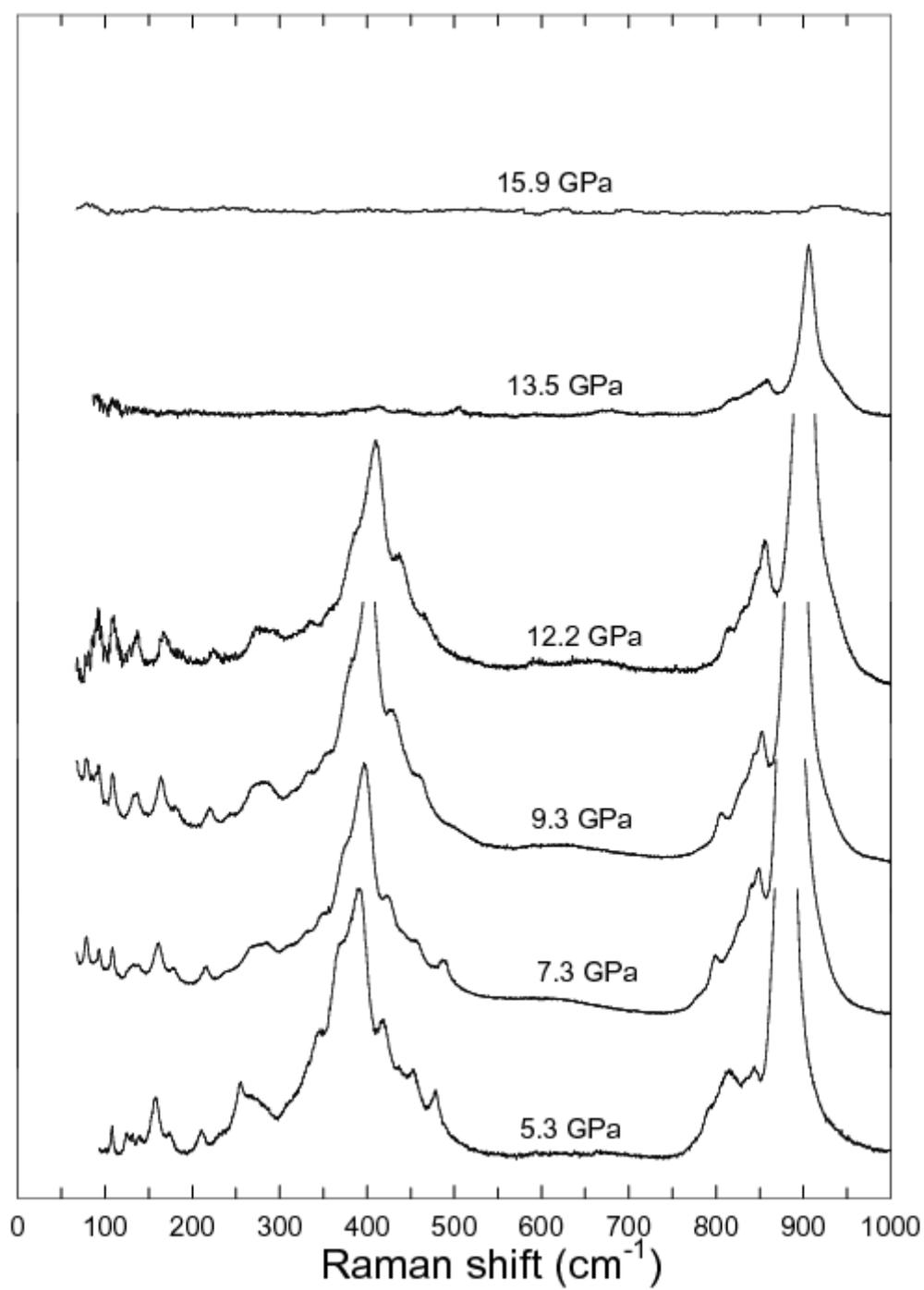

Figure 8(a)

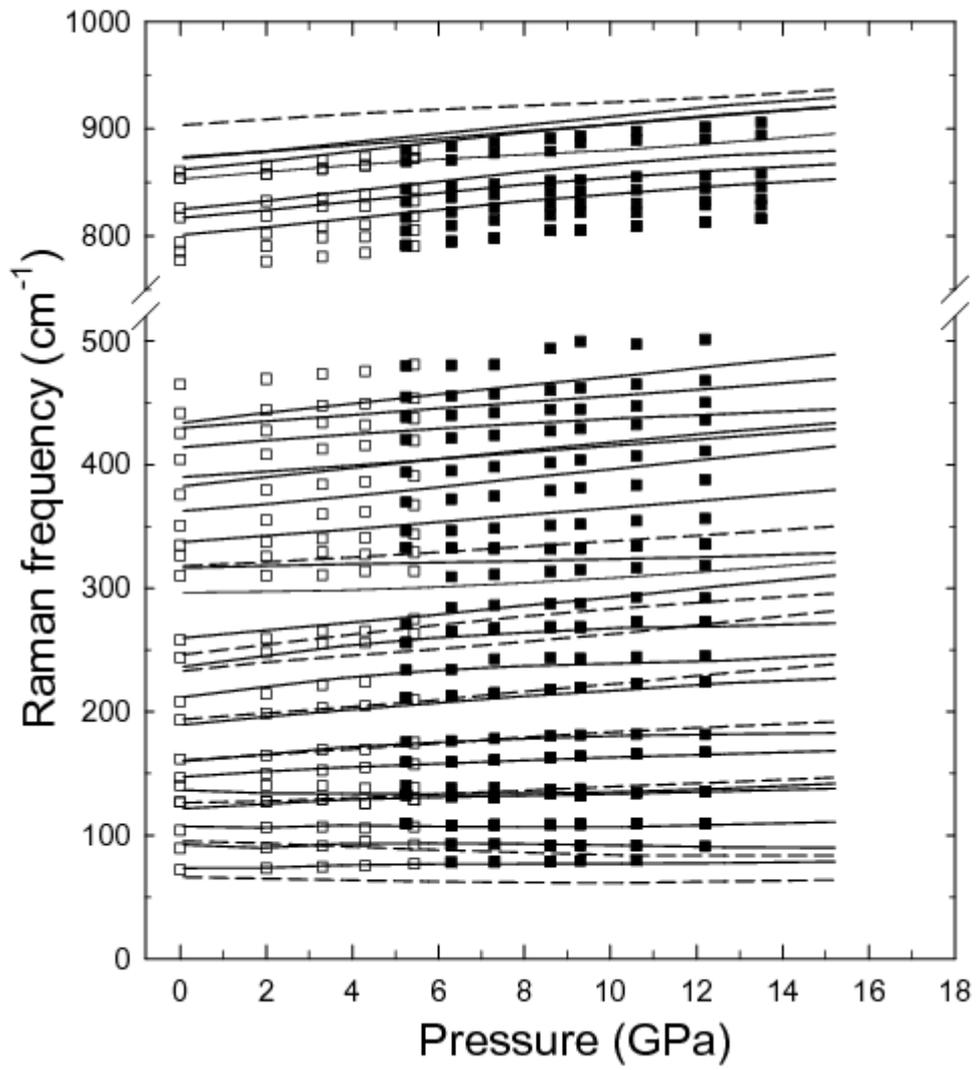

Figure 8(b)